\documentclass[prl,superscriptaddress,twocolumn]{revtex4}

\usepackage{graphicx}
\usepackage{psfrag}
\usepackage{amsmath}

\begin{document}
\newtheorem{theorem}{Theorem}
\newtheorem{acknowledgement}[theorem]{Acknowledgement}
\newtheorem{algorithm}[theorem]{Algorithm}
\newtheorem{axiom}[theorem]{Axiom}
\newtheorem{claim}[theorem]{Claim}
\newtheorem{conclusion}[theorem]{Conclusion}
\newtheorem{condition}[theorem]{Condition}
\newtheorem{conjecture}[theorem]{Conjecture}
\newtheorem{corollary}[theorem]{Corollary}
\newtheorem{criterion}[theorem]{Criterion}
\newtheorem{definition}[theorem]{Definition}
\newtheorem{example}[theorem]{Example}
\newtheorem{exercise}[theorem]{Exercise}
\newtheorem{lemma}[theorem]{Lemma}
\newtheorem{notation}[theorem]{Notation}
\newtheorem{problem}[theorem]{Problem}
\newtheorem{proposition}[theorem]{Proposition}
\newtheorem{remark}[theorem]{Remark}
\newtheorem{solution}[theorem]{Solution}
\newtheorem{summary}[theorem]{Summary}    
\def\r{{\bf{r}}}
\def\i{{\bf{i}}}
\def\j{{\bf{j}}}
\def\m{{\bf{m}}}
\def\k{{\bf{k}}}
\def\h{{\bf{h}}}
\def\kt{{\tilde{\k}}}
\def\qt{{\tilde{\q}}}
\def\mt{{\hat{t}}}
\def\mG{{\hat{G}}}
\def\mg{{\hat{g}}}
\def\mGa{{\hat{\Gamma}}}
\def\mS{{\hat{\Sigma}}}
\def\mT{{\hat{T}}}
\def\K{{\bf{K}}}
\def\P{{\bf{P}}}
\def\q{{\bf{q}}}
\def\Q{{\bf{Q}}}
\def\p{{\bf{p}}}
\def\x{{\bf{x}}}
\def\X{{\bf{X}}}
\def\Y{{\bf{Y}}}
\def\F{{\bf{F}}}
\def\R{{\bf{R}}}
\def\G{{\bf{G}}}
\def\bG{{\bar{G}}}
\def\mbG{{\hat{\bar{G}}}}
\def\M{{\bf{M}}}
\def\V{\cal V}
\def\tchi{\tilde{\chi}}
\def\tx{\tilde{\bf{x}}}
\def\tk{\tilde{\bf{k}}}
\def\tK{\tilde{\bf{K}}}
\def\tq{\tilde{\bf{q}}}
\def\tQ{\tilde{\bf{Q}}}
\def\si{\sigma}
\def\ep{\epsilon}
\def\hep{{\hat{\epsilon}}}
\def\al{\alpha}
\def\be{\beta}
\def\ep{\epsilon}
\def\bep{\bar{\epsilon}_\K}
\def\mep{\hat{\epsilon}}
\def\up{\uparrow}
\def\de{\delta}
\def\De{\Delta}
\def\up{\uparrow}
\def\dwn{\downarrow}
\def\ksi{\xi}
\def\etha{\eta}
\def\product{\prod}
\def\goto{\rightarrow}
\def\switch{\leftrightarrow}

\title{On the nature of superconductivity in the t-J model of
  high-T$_c$ superconductors} \author{Th.A.~Maier}
\affiliation{Computer Science and Mathematics Division, Oak Ridge
  National Laboratory, Oak Ridge, TN 37831-6114}

\date{\today}

\begin{abstract} 
  Using an extended dynamical cluster approximation we study
  superconductivity in the two-dimensional t-J model. In analogy to
  the extended dynamical mean field theory, non-local spin
  fluctuations are treated self-consistently with an additional
  Bosonic bath. To solve the effective cluster problem, we employ an
  extended non-crossing approximation which allows for transitions to
  the symmetry-broken state. At sufficiently low temperatures we find
  a stable superconducting solution with d-wave order parameter. Upon
  pairing, the exchange energy is lowered, consistent with an
  exchange-based pairing mechanism.
\end{abstract}

\maketitle

\paragraph{Introduction.} 
The nature of pairing in high-temperature superconductors (HTSC)
remains one of the most outstanding open problems in materials
science. After years of active research, we are still far from a
complete theoretical understanding of the rich physics observed in
these materials. Early in the history of HTSC it was realized that
electronic correlations play a crucial role not only for the
pairing-mechanism in HTSC but also for their unusual normal state
behavior \cite{anderson:htsc}.  Hence, models describing the behavior
of itinerant correlated electrons, in particular the Hubbard model and
its strong-coupling limit, the t-J model, were proposed to capture the
generic physics of HTSC \cite{anderson:hm,zhang:singlet}. However,
despite intensive studies, these models remain unsolved in two
dimensions.

Conventional (BCS) superconductors are well understood. Electrons can
lower their potential energy by forming Cooper-pairs. Chester
\cite{chester:ion} realized that this reduction in the electronic
potential energy actually corresponds to a decrease in the ionic
kinetic energy , thus providing a clear link between the pairing
mechanism and the lattice degrees of freedom. In analogy to Chester's
idea, Scalapino and White \cite{scalapino:cond} suggested an analogous
relation for the origin of the pairing energy in HTSC based on
numerical calculations for the t-J model. They proposed, if spin
correlations are involved in the pairing mechanism, that the
condensation energy in HTSC is provided by the exchange energy $J
\sum_{\langle ij\rangle}\langle \vec{S}_i\cdot \vec{S}_j\rangle$.
Consistent with this idea, neutron scattering experiments in HTSC have
shown the evolution of a low-energy spin-fluctuation resonance in the
superconducting state, which can be related to a reduction in exchange
energy \cite{demler:ns,dai:ns}.

This analogy to conventional BCS superconductors however, has recently
been contrasted by optical experiments in BSCCO \cite{marel:kineg}.
While in BCS superconductors the kinetic energy slightly increases
upon pairing, these experiments have shown that pairing in HTSC occurs
through a reduction of the electronic kinetic energy. In order to
unify these two seemingly controversial observations, it is important
to note that in the t-J model, magnetic exchange has mixed electronic
kinetic and Coulombic origins.  In addition, even the kinetic term in
the t-J Hamiltonian has Coulombic contributions due to the restriction
of no double-occupancy of a given site. Thus, while the Hubbard model
with its purely kinetic and Coulombic terms can be related to the
optical experiments, the t-J model has a well defined exchange energy
and hence is the model of choice to study the possibility of exchange
based pairing.

Using QMC simulations within the dynamical cluster approximation
(DCA), we have shown that pairing in the 2D Hubbard model indeed
happens through a reduction of the electronic kinetic energy
\cite{maier:kineg}, consistent with the experimental observation. To
explore the possibility of exchange-based pairing, we study in this
letter the nature of superconductivity in the 2D t-J model. Its
Hamiltonian reads in usual notation
\begin{equation}\label{eq:tjmodel}
H=-t \sum_{\langle ij\rangle,\sigma} \tilde{c}^\dagger_{i\sigma}\tilde{c}^{}_{j\sigma}+J\sum_{\langle ij\rangle} ( \vec{S}_i \cdot \vec{S}_j - \frac{1}{4} n_i n_j )\,,
\end{equation}
where the projected operators $\tilde{c}^{}_{i\sigma}$,
$\tilde{c}^\dagger_{i\sigma}$ act in the reduced Hilbert-space with
empty and doubly occupied sites only and $n_{i}=\sum_\sigma {\tilde
  c}^\dagger_{i\sigma}{\tilde c}^{}_{i\sigma}$ are their corresponding
number operators. Unless otherwise noted, we use the hopping integral
$t$ for the energy scale and set $J=0.3t$.

The possibility of $d_{x^2-y^2}$ pairing in the 2D t-J model was
indicated in a number of numerical studies of finite size systems (for
a review see \cite{dagotto:rmp}). Only recent numerical calculations
provided evidence for pairing in relatively large systems for
physically relevant values of $J/t$ \cite{sorella:tJ,scalapino:tJ}.
The question of whether pairing persists in the infinite size model
however remains open. Thus, simulations for the thermodynamic limit
are clearly desirable to further support the finite size results.

Simulations within the dynamical mean field theory (DMFT) take place
in the thermodynamic limit. An extended DMFT (EDMFT) study of the
pseudogap in the t-J model was presented in
Ref.~\cite{kroha:edmftletter}. But the lack of non-local correlations
in the DMFT method inhibits transitions to states with non-local order
parameters such as $d$-wave superconductivity.  The dynamical cluster
approximation (DCA) extends the DMFT by mapping the original lattice
model onto a periodic cluster with linear size $L_{c}$ embedded in a
self-consistent host.  As a result, dynamical correlations up to a
range $\sim L_{c}$ are treated accurately while the longer-ranged
physics is described on the mean-field level.

\paragraph{Method.}
A detailed discussion of the DCA for models with local interactions
like the Hubbard model was given in
Ref.~\cite{hettler:dca1,hettler:dca2}. The first Brillouin zone is
divided into $N_c=L_c^D$ cells where $D$ is the dimension, with each
cell represented by its center wave-vector, the cluster $\K$-points.
The reduction of the $N$-site lattice problem to an effective
$N_c$-site cluster problem is achieved by coarse-graining the compact
diagrams of the free energy, i.e. averaging over the $N/N_c$
wave-vectors $\kt$ within a cell.  Consequently, the Feynman diagrams
for the lattice self-energy collapse onto those of an effective
cluster model embedded in a Fermionic host which accounts for all
fluctuations arising from hopping of electrons between the cluster and
the rest of the system.

In the following we outline the modifications necessary to treat
models with non-local interactions like the t-J model. Since the
projected operators $\tilde{c}_{i\sigma}$ in the t-J Hamiltonian
(\ref{eq:tjmodel}) do not obey standard Fermionic commutation
relations, the diagrammatic derivation of the DCA cannot be applied to
the t-J model. However, the general idea of the DCA, i.e. the
approximation of irreducible lattice quantities by their corresponding
cluster quantities can still be applied to the t-J model. An extended
algorithm for models, such as the t-J model, with extended range
interactions was formulated in Ref.~\cite{hettler:dca2} for the DCA.
For the cluster size $N_{c}=1$ this algorithm corresponds to the
extended DMFA (EDMFA) \cite{si:edmft1}. Fluctuations associated with
the non-local interactions are accounted for by an additional Bosonic
bath which reflects the influence of the host on the two-particle
level. In the case of the t-J model, it represents the fluctuating
magnetic fields induced by $J$. 

In the DCA, the effective cluster
model is conveniently represented in reciprocal cluster space
\begin{eqnarray}
  \label{eq:clmod}
  H_{cl}&=&\sum_{\K,\sigma}\bar{\epsilon}_\K \tilde{c}^\dagger_{\K\sigma}
           \tilde{c}^{}_{\K\sigma} + \sum_\Q \left[\bar{J}_\Q {\bf
           S}_\Q\cdot {\bf S}_{-\Q}-\frac{1}{4}
           \rho_\Q\rho_{-\Q}\right]\nonumber\\
        &+&\sum_{\K,\tk,\sigma}\left[V_{\K,\kt}\tilde{c}^\dagger_{\K\sigma} 
           a^{}_{\K+\kt} +h.c.\right] \nonumber\\
        &+&\sum_{\Q,\qt}\left[ I_{\Q,\qt} {\bf S}_\Q\cdot({\bf \Phi}_{\Q+\qt} +
        {\bf \Phi}^\dagger_{-\Q-\qt}) +h.c.\right]\nonumber\\
        &+& \sum_{\k\sigma}\lambda_\k
        a^\dagger_{\k\sigma}a^{}_{\k\sigma} + \sum_\q \omega_\q {\bf
        \Phi}^\dagger_\q \cdot {\bf \Phi}^{}_\q\,. 
\end{eqnarray}
The operators $\tilde{c}^{(\dagger)}_{\K\sigma}$ create (destroy)
Fermions on the cluster with wave-vector $\K$ and ${\bf S}_\Q$
($\rho_\Q$) are their corresponding spin- (density-) operators. The
sums over $\kt$ and $\qt$ run over the momenta within a DCA cell about
the cluster momenta $\K$ and $\Q$. In the DCA, non-local parameters
are coarse-grained, i.e.  $\bar{\epsilon}_\K=N_c/N\sum_\kt
\epsilon_{\K+\kt}$ and $\bar{J}_\Q=N_c/N \sum_\qt J_{\Q+\qt}$ are
given by the cell-averages of the lattice Fourier-transforms
$\epsilon_\k$ and $J_\q$ of $t$ and $J$.  The Fermionic and Bosonic
hosts are described by the auxiliary operators
$a^{(\dagger)}_{\k\sigma}$ and ${\bf \Phi}^{(\dagger)}_\q$ with
corresponding energies $\lambda_\k$ and $\omega_\q$. When the system
becomes superconducting, an additional U(1) symmetry breaking term
\begin{equation}
  \label{eq:U1}
  H_{SC}=\sum_\k \left[ \Delta_\k
  a^\dagger_{\k\downarrow}a^\dagger_{-\k\uparrow} + h.c. \right]
\end{equation}
has to be introduced to simulate the existence of paired states in the
host. 

The solution of the cluster problem given by $H_{cl}+H_{SC}$
yields the cluster Green function ${\bf G}_c(\K,\omega)$, cluster spin
susceptibility $\chi_c(\Q,\omega)$ and their corresponding
self-energies ${\bf \Sigma}(\K,\omega)$ and $M(\Q,\omega)$. In the
DCA, lattice self-energies are approximated by their corresponding
cluster quantities, i.e. the lattice Green function is written as
\begin{equation}
  \label{eq:lGf}
  {\bf G}(\k,\omega)=\left[ \omega\tau_{o}-\epsilon_{\K+\kt}\tau_{3}-{\bf
\Sigma}({\bf K},\omega) \right]^{-1}\,.
\end{equation}
To allow for the possibility of superconducting order we used the
Nambu-Gorkov matrix representation \cite{schrieffer:book} with
Pauli-spin matrices $\tau_i$. The 11-component of ${\bf G}$ is given
by the normal single-electron Green function $G(\k,z)=\langle\langle
c^{}_{\k\uparrow}; c^\dagger_{\k\uparrow} \rangle\rangle_z$, while the
12-component is the anomalous Green function $F(\k,z)= \langle\langle
c^{}_{\k\uparrow}; c^{}_{-\k\downarrow} \rangle\rangle_z$ and only
finite in the superconducting state. The lattice spin susceptibility
becomes
\begin{equation}
  \label{eq:lss}
  \chi(\q,\omega)=\frac{1}{M(\Q,\omega)+J_\q}\,.
\end{equation}
Self-consistency is imposed by demanding that the cluster Green
function ${\bf G}_c$ and spin-susceptibility $\chi_c$ are equal to the
coarse grained, i.e. cell averaged lattice Green function
\begin{eqnarray}\label{eq:cgg}
\bar{{\bf G}}({\bf K},z)&=&\frac{N_{c}}{N}\sum\limits_\kt {\bf
G}(\K+\kt,\omega)\nonumber\\
&=&\left[ \omega\tau_{o}-\bar{\epsilon}_{{\bf K}}\tau_{3}-{\bf
\Sigma}({\bf K},\omega)-{\bf \Gamma}({\bf K},\omega)
\right]^{-1}
\end{eqnarray}
and lattice spin-susceptibility
\begin{eqnarray}\label{eq:cgc}
\bar{\chi}({\bf Q},\omega)&=& \frac{N_c}{N}\sum_\qt \chi(\Q+\qt,\omega)\nonumber\\
&=&\frac{1}{M({\bf Q},\omega)+\bar{J}(\Q)+\Theta({\bf Q},\omega)}
\end{eqnarray}
respectively, where the sums run over momenta within the DCA cell about the cluster momentum $\K$. The DCA self-consistent problem is then solved by
iteration. In this procedure, the auxiliary parameters of the cluster
model $\lambda_\k$, $\omega_\q$, $\Delta_\k$, $V_{\K,\kt}$ and
$I_{\Q,\qt}$ are replaced by the hybridization functions ${\bf
  \Gamma}(\K,\omega)$ and $\Theta(\Q,\omega)$ which are
self-consistently determined from the coarse grained lattice
quantities $\bar{\bf G}$, Eq.~(\ref{eq:cgg}) and $\bar{\chi}$,
Eq.~(\ref{eq:cgc}). Hence, the particular form of the auxiliary
degrees of freedom in Eqs.~(\ref{eq:clmod}) and (\ref{eq:U1}) is
irrelevant and thus does not impose any further approximation.

Although the complexity of the problem is highly reduced by the DCA,
the solution of the effective quantum cluster problem remains
non-trivial. The additional coupling to the Bosonic bath renders many
traditional methods used to solve the purely Fermionic cluster model
like e.q. QMC difficult to apply. Here we employ an extended version
of the non-crossing approximation which is based on an infinite
resummation of certain classes of diagrams in a pertubational
expansion in the couplings to the Fermionic and Bosonic baths. A
detailed derivation of the NCA equations for the EDMFT quantum
impurity model ($N_c=1$) was given in Ref.\cite{kroha:edmft}. The
application of the NCA to the Fermionic DCA cluster problem was
discussed in Ref.\cite{maier:dca1} and the modifications necessary to
treat the superconducting state in Ref.\cite{maier:diss}. A detailed
discussion of the NCA equations for the extended cluster problem
embedded in the additional Bosonic bath will be presented elsewhere.

\paragraph{Results.}
We performed simulations for the cluster size $N_c=4$, the smallest
possible cluster size that can capture pairing with the $d$-wave order
parameter in the t-J model. Thus, this theory should be regarded as
the mean-field theory for $d$-wave superconductivity in the t-J model.
In particular, we choose the conventional set of cluster $\K$-points
$\{ (0,0), (\pi,0), (0,\pi), (\pi,\pi)\}$.

\begin{figure}[htbp]
  \centering \psfrag{DOS}[][][0.8]{$N(\omega)\times t$}
  \psfrag{omega}[][][0.8]{$\omega/t$} 
  \psfrag{doping}[][][0.8]{$\delta$} 
  \psfrag{Tc}[][][0.8]{$T_c/t$} 
  \psfrag{coarse-grained anomalous Green function}[][][0.8]{$\Re e {\bar F}(\K,\omega)\times t$} 
  \psfrag{K1}[c][][0.6]{$(\pi,0)$}
  \psfrag{K2}[c][][0.6]{$(0,\pi)$}
  \includegraphics*[height=5cm]{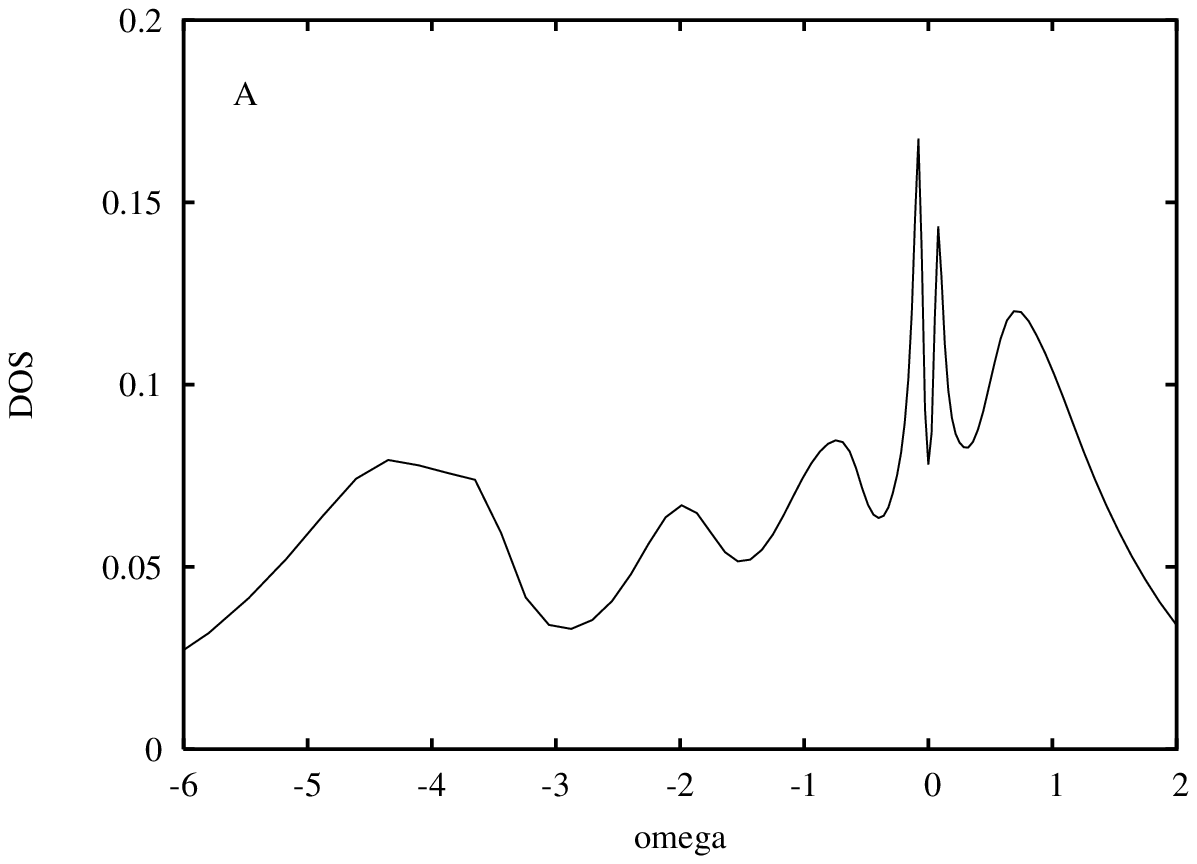}
  \includegraphics*[height=3.5cm]{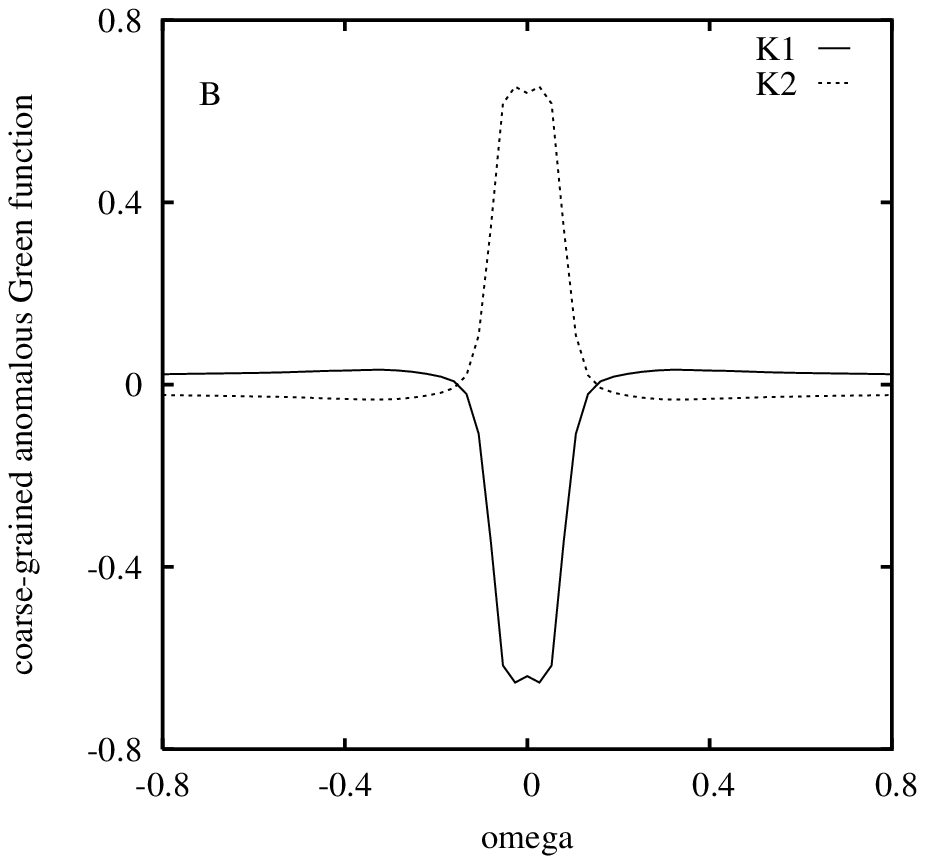}\includegraphics*[height=3.5cm]{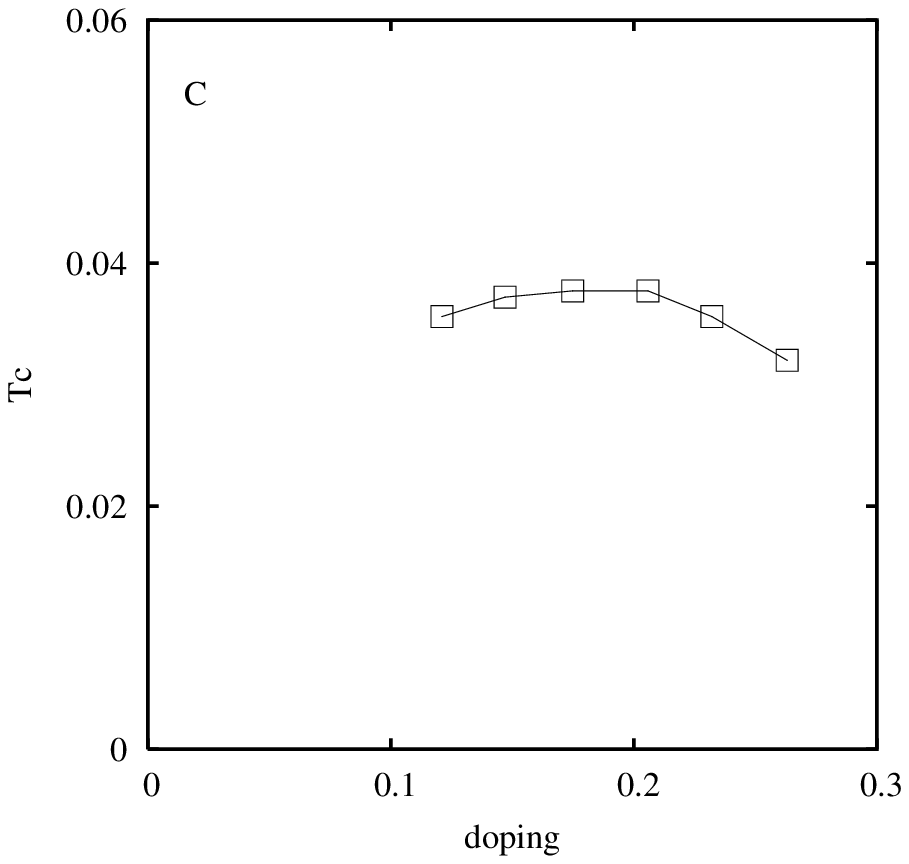}
  \caption{(A) Density of states and (B) real part of the coarse-grained
  anomalous Green function in the superconducting state of the
  2D t-J model at 20\% doping at $T=0.0348t$. (C) Doping dependence of the critical temperature $T_c$.}
  \label{fig:0}
\end{figure}

To show that our technique is indeed able to describe
superconductivity within the t-J model, we present in Fig.~\ref{fig:0}
our results for the single-particle spectrum at 20\% doping
($\delta=0.20$) and temperature $T=0.0348t$ below the superconducting
phase transition at $T_c=0.0377t$.  The density of states $N(\omega)$
shown in Fig.~\ref{fig:0}A develops a pseudogap at the chemical
potential ($\omega=0$), when the anomalous Green-function
$\bar{F}(\K,\omega)=\langle\langle \tilde{c}_{\K\uparrow};
\tilde{c}_{-\K\downarrow} \rangle\rangle$ shown in Fig.~\ref{fig:0}B,
becomes finite. The $\K$-dependence of the coarse-grained anomalous
Green function, i.e. $\bar{F}((\pi,0),\omega) =
-\bar{F}((0,\pi),\omega)$ and zero otherwise (not shown), is
consistent with a $d_{x^2-y^2}$-symmetry of the order parameter.
Fig.~\ref{fig:0}C shows the doping dependence of the transition
temperature. $T_c(\delta)$ has a maximum $T_c^{max}\approx 0.0377
t\approx 110{\rm K}$ (for $t=0.25{\rm eV}$) at the optimal doping
$\delta\approx 0.20$ and decreases with both increasing and decreasing
doping. Due to the breakdown of the NCA technique at very low
temperatures, the phase-diagram cannot be extended beyond the region
shown and in particular no prediction can be made for the critical
doping beyond which superconductivity vanishes.

It is interesting to
note that without the additional coupling to the Bosonic host we were
not able to stabilize a superconducting solution, at least not above
the lowest temperatures we can simulate with our technique. This shows
the importance of a self-consistent treatment of the spin-fluctuations
between the cluster and host.

To explore the nature of pairing in our simulations, we calculate the
exchange energy $E_{xc}$ from the spin-susceptibility
$\chi(\q,\omega)$, Eq.~(\ref{eq:lss}), via the fluctuation-dissipation
theorem as
\begin{eqnarray}
\label{eq:Exc}
E_{xc} &=& \frac{1}{N}\sum_\q J(\q) \langle \vec{S}(\q)\cdot \vec{S}(-\q)\rangle\nonumber\\
&=&-\frac{3}{\pi}\frac{1}{N}\sum_\q J(\q) \int d\omega \frac{\Im m \chi(\q,\omega)}{1-e^{-\beta\omega}}\,.
\end{eqnarray} 
To obtain results for the normal state below the transition
temperature $T_c$ we suppress superconductivity by not allowing for a
finite order parameter in our simulation.

\begin{figure}[htbp]
  \centering
  \psfrag{Exc}[][][0.8]{$E_{xc}/t$}
  \psfrag{T}[][][0.8]{$T/t$}  
  \includegraphics*[height=5cm]{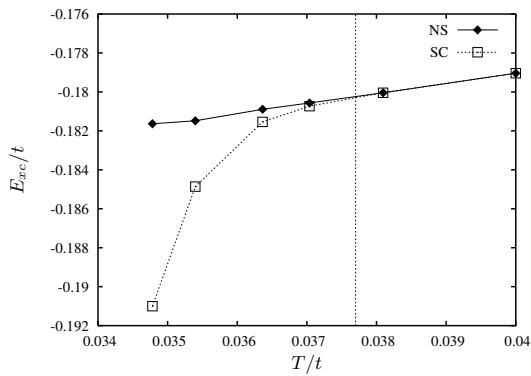}
  \caption{Exchange energy as a function of temperature at 20\% doping across the superconducting transition in the normal (NS) and superconducting (SC) states.}
  \label{fig:1}
\end{figure}

The exchange energies in the normal and superconducting states as a
function of temperature close to the critical temperature $T_c$ at
20\% doping are shown in Fig.~\ref{fig:1}.
Our results for the nature of pairing in the 2D t-J model are
consistent with an exchange-based pairing mechanism: Below $T_c$, the
exchange energy in the superconducting state is reduced compared to
the normal state, while the energy coming from the kinetic (first)
term in the t-J model, Eq.~(\ref{eq:tjmodel}), slightly increases upon
pairing (not shown). It is important to remember that in the mapping
of the Hubbard model to the t-J model, the exchange interaction $J$
comes from virtual hopping processes between neighboring sites and
thus has kinetic character. Hence the present results are consistent
with our results for the Hubbard model where we found the
superconducting instability to be accompanied by a lowering of the
electronic kinetic energy \cite{maier:kineg}. The nature of pairing in
the t-J model however is more similar to that in BCS superconductors:
The energy of the exchange coupling which plays a similar role to that
of the lattice degrees of freedom in BCS superconductors, is reduced.

\begin{figure}[htbp]
  \centering
  \psfrag{omega}[][][0.8]{$\omega/t$}
  \psfrag{Im(coarse-grained spin-susceptibility)}[][][0.8]{$\Im m {\bar \chi}(\Q,\omega)$}  
 \includegraphics*[height=5cm]{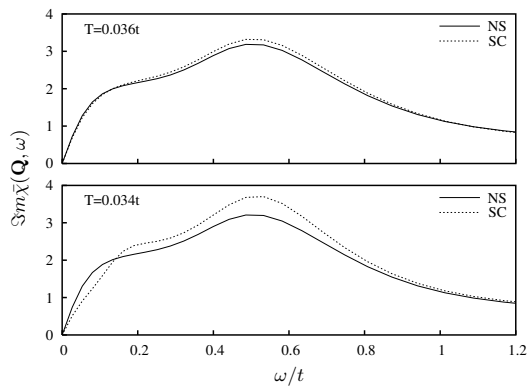} 
  \caption{Imaginary part of the coarse-grained dynamic spin-susceptibility in the 2D t-J model for $\Q=(\pi,\pi)$ at T=0.036t (top) and T=0.034t (bottom) in the superconducting (SC) and normal (NS) states.}
  \label{fig:2}
\end{figure}

Neutron scattering experiments relate the gain in exchange energy to
the development of a resonance in the spin-susceptibility near
$\Q=(\pi,\pi)$ and $40\, {\rm meV}$. According to Eq.~(\ref{eq:Exc}),
the exchange energy $E_{xc}$ is reduced when the spectral weight in
the spin-susceptibility $\chi(\q,\omega)$ is increased near
$\q=(\pi,\pi)$ since $J(\q)=J(\cos q_x + \cos q_y)<0$ in this region
in $\q$-space. In Fig.~\ref{fig:2} we plot the imaginary part of the
dynamic coarse-grained spin-susceptibility at $\Q=(\pi,\pi)$ in the
normal and superconducting states at 20\% doping for two different
temperatures below the critical temperature $T_c$. This $\q$-space
averaged quantity allows us to infer the changes in spectral weight
which according to Eq.~(\ref{eq:Exc}) reduce the exchange energy.
Although our data does not display the development of the narrow
resonance peak at $\sim 40\, {\rm meV}$ observed in experiments, we
find that the reduction in exchange energy indeed originates in an
increase of spectral weight in $\chi_s(\q,\omega)$ in the region near
$\Q=(\pi,\pi)$ consistent with experiments. 

To summarize, we have presented extended DCA/NCA simulations for the
2D t-J model which show evidence for a $d$-wave superconducting state
in the thermodynamic limit. We find that the pairing instability is
accompanied by a reduction of the exchange energy consistent with
exchange-based pairing. In agreement with neutron scattering
experiments, this exchange energy reduction originates in an increase
in the magnetic response around the antiferromagnetic wave-vector
$\Q=(\pi,\pi)$.

\paragraph*{Acknowledgements} 
It is a pleasure to acknowledge useful discussions with M. Jarrell, T.
Pruschke and M. Vojta. Research performed as a Eugene P. Wigner Fellow
and staff member at the Oak Ridge National Laboratory, managed by
UT-Battelle, LLC, for the U.S. Department of Energy under Contract
DE-AC05-00OR22725.

\bibliography{mybib}

\end{document}